\documentstyle[12pt,lathuile,epsfig]{article}
\newcommand{\gsp}{\mbox{$\gamma^* p$}}

\newcommand{\PO}{{\rm l \! P }}
\newcommand{\rh}{\mbox{$\varrho$}}

\newcommand{\ph}{\mbox{$\phi$}}
\newcommand{\om}{\mbox{$\omega$}}
\newcommand{\jpsi}{\mbox{$J/\psi$}}

\newcommand{\qsq}{\mbox{$Q^2$}}
\newcommand{\qsqz}{\mbox{$Q^2_0$}}
\newcommand{\W}{\mbox{$W$}}

\newcommand{\ttr}{\mbox{$t$}}

\newcommand{\xpom}{x_{\PO}}
\newcommand{\alpom}{\mbox{$\alpha_{\PO}$}}

\newcommand{\pom}{{I\!\!P}}

\newcommand{\fdv}{\mbox{$F_2^{D(3)}$($Q^2$, $\xpom$, $\beta$)}}
\newcommand{\fdtv}{\mbox{$F_2^{D(2)}$($Q^2$, $\beta$)}}

\newcommand{\gevsq}{\mbox{${\rm GeV}^2$}}

\def\Journal#1#2#3#4{{#1} {\bf #2} (#3) #4}

\def\PLB{Phys. Lett.   {\bf B}}
\def\PRL{Phys. Rev. Lett.}
\def\PRD{Phys. Rev. {\bf D}}
\def\ZPC{Z. Phys. {\bf C}}
\def\EJC{Eur. Phys. J. {\bf C}}

\def\jetp#1#2#3 {JETP Lett. {\bf#1} (#2) #3}
\begin{document}
\title{ 
RESULTS ON DIFFRACTION AT HERA AND TEVATRON 
}
\author{
Barbara Clerbaux  \\
{\em Inter-University Institute for High Energies (IIHE), Brussels,
     Belgium } \\
{\em (Now at CERN, Geneva, Switzerland)} \\ [2mm]
For the H1, ZEUS, CDF and D0 Collaborations
}
\maketitle
\baselineskip=14.5pt
\begin{abstract}
Recent results on diffraction at HERA, from the H1 and ZEUS Collaborations,
and at Tevatron, from the CDF and D0 Collaborations, are reviewed.
The measurement of the diffractive structure function at HERA is presented,
and the pomeron structure function is extracted from QCD fits. Diffractive 
dijet production is given as an example analysis of the 
diffractive final state. These analysis are consistent with a leading 
gluon partonic structure for the pomeron.
Hard single diffraction, double diffraction and double pomeron exchange
processes are studied at the Tevatron and complement the studies performed at
HERA. Finally, HERA results on the diffractive (exclusive) production of a vector particle,
either a vector meson or a photon, are presented and compared to models based on
perturbative QCD.
\end{abstract}
\baselineskip=17pt
\newpage
%
\section{Introduction}

The understanding of diffractive interactions is of fundamental 
importance as they govern the high energy behaviour of the elastic 
cross sections and thus of the total cross sections (via the optical theorem).

In the 70's, the diffractive processes were intensively studied in
hadron--hadron interactions and were well described by the Regge
phenomenology. In this framework, the elastic scattering is attributed
at high energy to the exchange between the incoming hadrons of a 
colourless object, the pomeron. The energy dependence of the elastic
cross section is parameterised as
$ds/dt \propto s^{2(\alpom (t) -1)}$, and 
depends on the trajectory of the pomeron $\alpom (t)$:
$\alpom (t) = 1.08 + 0.25 t$,
$t$ being the square of the four-momentum transfer.
The total, elastic, and diffractive cross sections,
thus exhibit a ``soft" energy dependence, at high energy.

An important result of HERA studies is that,
in contrast to the slow increase with energy of hadron-hadron cross sections
(``soft" behaviour), the total $\gamma^* p$ cross section has
a strong (``hard") energy dependence in the deep inelastic scattering
(DIS) domain, which is attributed to a
fast rise with energy of the gluon density in the proton.
The QCD pomeron being described as a gluonic system,
a ``hard" behaviour is thus also expected in diffractive interactions.

The interest is now to understand the diffractive interaction in
the framework of the QCD theory, and in particular, to study the partonic 
structure of the pomeron. 
%
\section{Diffractive structure function at HERA}

\begin{figure}[t]
\begin{center}
\setlength{\unitlength}{1cm}
  \begin{picture}(12.0,3.8)
   \put(-1.0,0.0){\epsfig{figure=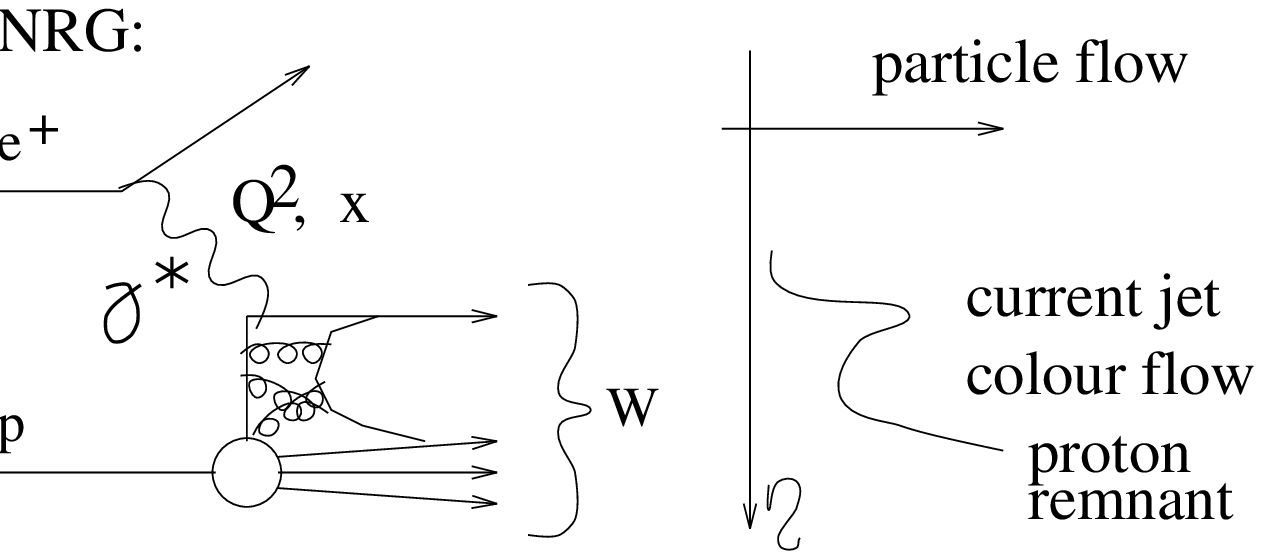,height=3.8cm,width=5.5cm}}
   \put(6.8,0.0){\epsfig{figure=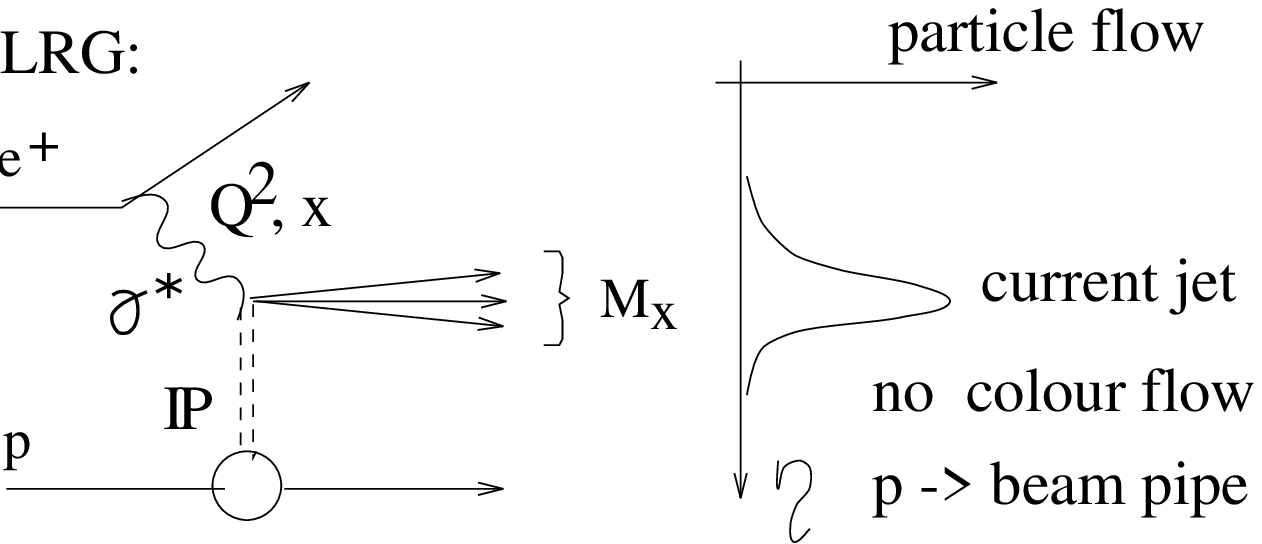,height=3.8cm,width=5.5cm}}
   \put(2.0,-0.6){(a)}
   \put(9.0,-0.6){(b)}
  \end{picture}
\end{center}
\vspace*{-0.1cm}
\caption{a) Deep Inelastic Scattering, and b) diffractive interaction.
\label{fig:diag}}
\end{figure}

Diffractive interactions at HERA account for above 10 \% of the
deep inelastic scattering (DIS) events. 
The diffractive cross section is measured by selecting the
events: $e+p \rightarrow e+X+Y$, where the two hadronic systems $X$ and $Y$ are separated 
by a large rapidity gap, devoid of particles, $Y$ being the system closest to the 
outgoing proton beam direction. The topology of diffractive events at HERA is presented
in Fig.~\ref{fig:diag}b, in contrast to the topology of 
events in the DIS regime (Fig.~\ref{fig:diag}a).
The measurement of the cross section is given in term of the diffractive structure
function \fdv. The variable
\qsq~is the negative of the square of {\it q},
the four--momentum carried by the virtual photon, and
\begin{equation}
\xpom= q.(-k)/(q.p) \approx \frac{\qsq+M_X^2}{\qsq+W^2};~~
\beta=\qsq /(2q.(-k)) \approx \frac{\qsq}{\qsq+M_X^2},
                                           \nonumber
\end{equation}
where {\it k} ($p$) is the four--momentum carried by the pomeron (incident proton),
$M_X$ is the invariant mass of the dissociative photon system, and \W~
is the energy in the \gsp~center of mass system.
For diffractive interactions, the kinematical variable~\ttr~= $k^2$
is expected to be small.
\fdv~is integrated over \ttr.
In a picture of diffractive interactions where a pomeron is emitted
from the proton, the pomeron having a partonic structure, $\xpom$ is the
fraction of the proton momentum carried by the pomeron, and $\beta$ is
the fraction of the pomeron momentum carried by the quark interacting
with the virtual photon. The product $\xpom \cdot \beta$ = $x$, the
usual Bjorken scaling variable.

It has been shown that the amplitudes for diffractive deep inelastic scattering
factorise out into a 
part which depends on $\xpom$ (a 'pomeron flux factor') and a structure function
$F_2^D (\beta, \qsq)$ corresponding to a universal partonic structure of diffraction:
$F_2^{D(3)}\,(x_{I\!\!P},\,\beta,\,Q^2) \propto f(\xpom) \cdot  F_2^{D}\,(\beta,\,Q^2)$.

\begin{figure}[tbh]
\begin{center}
\setlength{\unitlength}{1cm}
  \begin{picture}(7.0,6.0)
 \epsfig{file=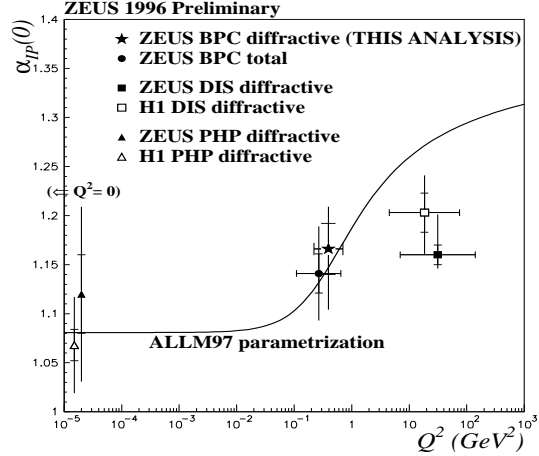,width=7.cm,height=6.cm}
  \end{picture}
\end{center}
\vspace*{-0.5cm}
\caption{HERA Measurements~\protect\cite{f2d3ZEUSlowq2} of the pomeron intercept
        $\alpha_\pom(0)$ as a function of \qsq.}
\label{fig:alpha}
\end{figure}

In a Regge approach, the pomeron flux factor follows a power law:
$f(\xpom)  \propto (1/\xpom)^{2 \alpha_{\pom} -1}$.
From the measurement of the $\xpom$ dependence of the diffractive structure 
function, the pomeron intercept $\alpha_\pom(0)$ can be extracted for 
different \qsq\ values~\cite{f2d3ZEUSlowq2}.
As is presented in Fig.~\ref{fig:alpha}, the pomeron intercept for \qsq\ $>$ 0
has a higher value than 1.08, which is typical of hadron--hadron interactions.
The transition from a soft to hard behaviour happens at
low \qsq\ values.

\begin{figure}[p]
\begin{center}
\setlength{\unitlength}{1cm}
  \begin{picture}(12.,7.0)
    \put(0.0,0.0){\epsfig{file=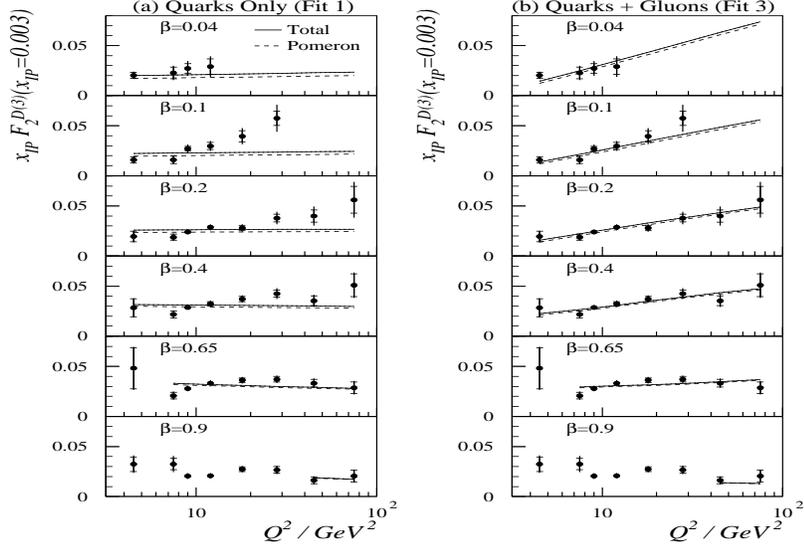,width=12.0cm,height=8.0cm}}
    \put(0.0,-23.0){\epsfig{file=whitebox.eps,width=12.0cm,height=23.0cm}}
   \end{picture}
\end{center}
\vspace*{-0.5cm}
\caption{H1 measurements~\protect\cite{h1_f2d3} of $\xpom \cdot F_2^{D(3)}$ 
  extrapolated to $\xpom$ =0.003, as
  a function of \qsq, for different $\beta$ bins. The curves represent the
  DGLAP QCD evolution of the (\qsq, $\beta$) dependence of
  \fdtv, assuming at the starting scale of \qsqz~= 3.0 \gevsq~a)
  quarks only, b) both quarks and gluons.
\label{fig:strqcd}}
\end{figure}

\begin{figure}[p]
\begin{center}
\setlength{\unitlength}{1cm}
  \begin{picture}(9.0,7.0)
   \put(0.0,0.0){\epsfig{file=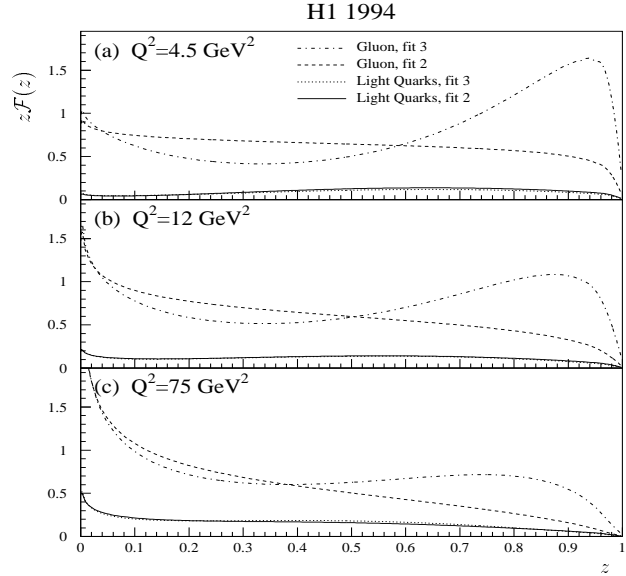,%
        width=9.0cm,height=8.0cm}}
  \end{picture}
\end{center}
\vspace*{-0.5cm}
\caption{The sum of the light quark distributions and the gluon distribution
   for 2 possible fits:
   for fit 2 ('flat' gluon) and fit 3 ('peaked' gluon), see~\protect\cite{h1_f2d3}, 
   shown for (a) \qsq\ = 4.5 \gevsq, (b)  \qsq\ = 12 \gevsq\
   and (c) \qsq\ = 75 \gevsq.}
\label{fig:gluon}
\end{figure}

The structure function $F_2 (\beta, \qsq)$, multiplied by $\xpom$ for clarity,
is presented in Fig.~\ref{fig:strqcd} for a fixed value of $\xpom$ =0.003, as a
function of \qsq, for different
$\beta$ bins~\cite{h1_f2d3}.
It differs markedly from the proton structure
function: \fdtv~is still large at large $\beta$,
and the rise with \qsq~persists for values of $\beta$ much larger than
$\approx$ 0.15. This is an indication for a dominant
gluonic component in the pomeron.  A QCD fit was performed on the data, using the
Altarelli--Parisi (DGLAP) evolution equations applied to the 
parton distributions defined, as a function of beta,
for a starting scale $Q_0$. Two different initial
parton distributions were considered for
\qsqz~= 3.0 \gevsq:
only quarks, and both quarks and gluons.
In the first case (see Fig.~\ref{fig:strqcd}a), the DGLAP evolution
fails to reproduce rising scaling
violations at large $\beta$, whereas the mixed quark and gluon case (see
Fig.~\ref{fig:strqcd}b) can describe this rise.
For the latter parameterisation, the parton distributions are dominated,
throughout the \qsq~range, by gluons, which carry a large fraction of the 
pomeron momentum; this fraction decreases only slowly at large $\beta$
as \qsq~increases (see Fig.~\ref{fig:gluon}). 


\begin{figure}[bth]
\begin{center}
\setlength{\unitlength}{1cm}
  \begin{picture}(12.0,6.5)
   \put(-0.4,1.0){\epsfig{file=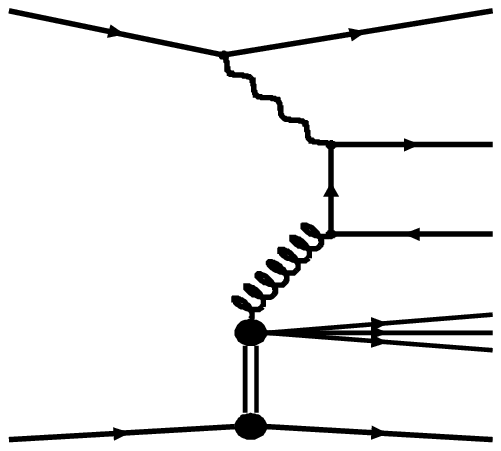,%
        width=4.0cm,height=4.0cm}}
   \put(5.0,0.0){\epsfig{file=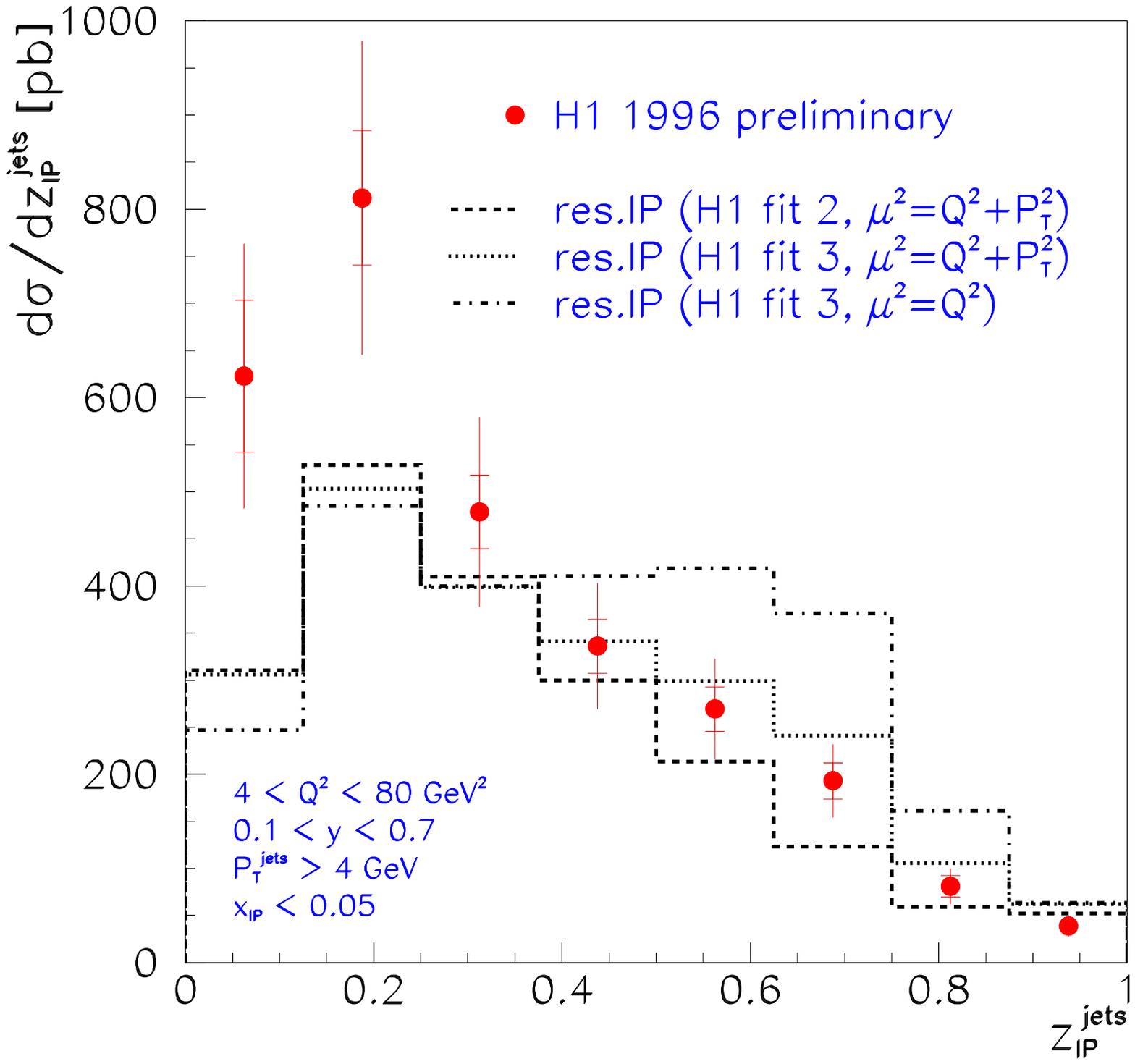,%
        width=7.0cm,height=6.5cm}}
   \put(1.7,-0.2){(a)}
   \put(8.5,-0.2){(b)}
   \put(-0.9,1.0){\large $p$}
   \put(-0.9,5.0){\large $e$}
   \put(4.0,1.0){\large $p$}
   \put(4.0,5.0){\large $e$}
   \put(4.0,2.8){\large $q$}
   \put(4.0,3.6){\large $\bar q$}
   \put(1.3,2.6){\large $g$}
   \put(1.2,3.8){\large $\gamma$}
   \put(0.8,1.5){\large $\xpom$}
  \end{picture}
\caption{(a) Dominating leading order QCD process, in which a $q\bar q$ pair
   is produced via photon gluon fusion;
   (b) H1 measurement of the differential cross section for diffractive dijet
   production, as a function of $z_\pom$, the momentum fraction of the pomeron
   carried by the parton
   entering the hard scattering process~\protect\cite{h1_dijet}.
   The data are shown together with predictions from a partonic pomeron model, 
   with 'flat' (fit 2) and 'peaked' (fit 3)
   gluon dominated parton densities for the pomeron, evolved at two different QCD scales.}
\label{fig:finstate}
\end{center}
\end{figure}

To complete the understanding of diffraction, studies of the diffractive final
state: jet production, hadron transverse momentum distribution,
hadron energy flow, charmed particle production... are performed at HERA.
The pomeron structure functions extracted from QCD fits to 
inclusive diffractive DIS are convoluted with scattering amplitudes,
to describe the specific final states.
The analysis of these different final states is consistent with the
picture of a leading gluonic component in the pomeron.
This supports the idea of universality of parton distributions in the pomeron.
As an example, the differential
cross section measurement for diffractive dijet electroproduction 
(4 $<$ \qsq\ $<$ 80 \gevsq) with $p_T^{jet} > 4$ GeV, is presented in
Fig.~\ref{fig:finstate}b. This process is a direct probe of the gluon
in the pomeron, via the boson gluon fusion mechanism (see Fig~\ref{fig:finstate}a). 
Figure~\ref{fig:finstate}b
shows the distribution of the variable 
$z_\pom = (M^2_{JJ}+\qsq)/(M_X^2 + \qsq)$ (where $M_{JJ}$ is the invariant mass
of the two jet system), which represents the fraction of the pomeron momentum carried by
the partons entering the hard process. The bulk of events have $z_\pom$ $<$ 1, i.e. 
$M_{JJ}$ $<$ $M_X$, indicating that pomeron remnants carry a
significant fraction of the pomeron momentum. 
The predictions using the QCD fit to the diffractive structure function, are in good agreement,
in shape and in normalisation, with the data.

%
\section{Hard diffraction at Tevatron}

\begin{figure}[t]
\begin{center}
\setlength{\unitlength}{1cm}
  \begin{picture}(15.0,6.5)
   \put(0.0,-0.5){\epsfig{file=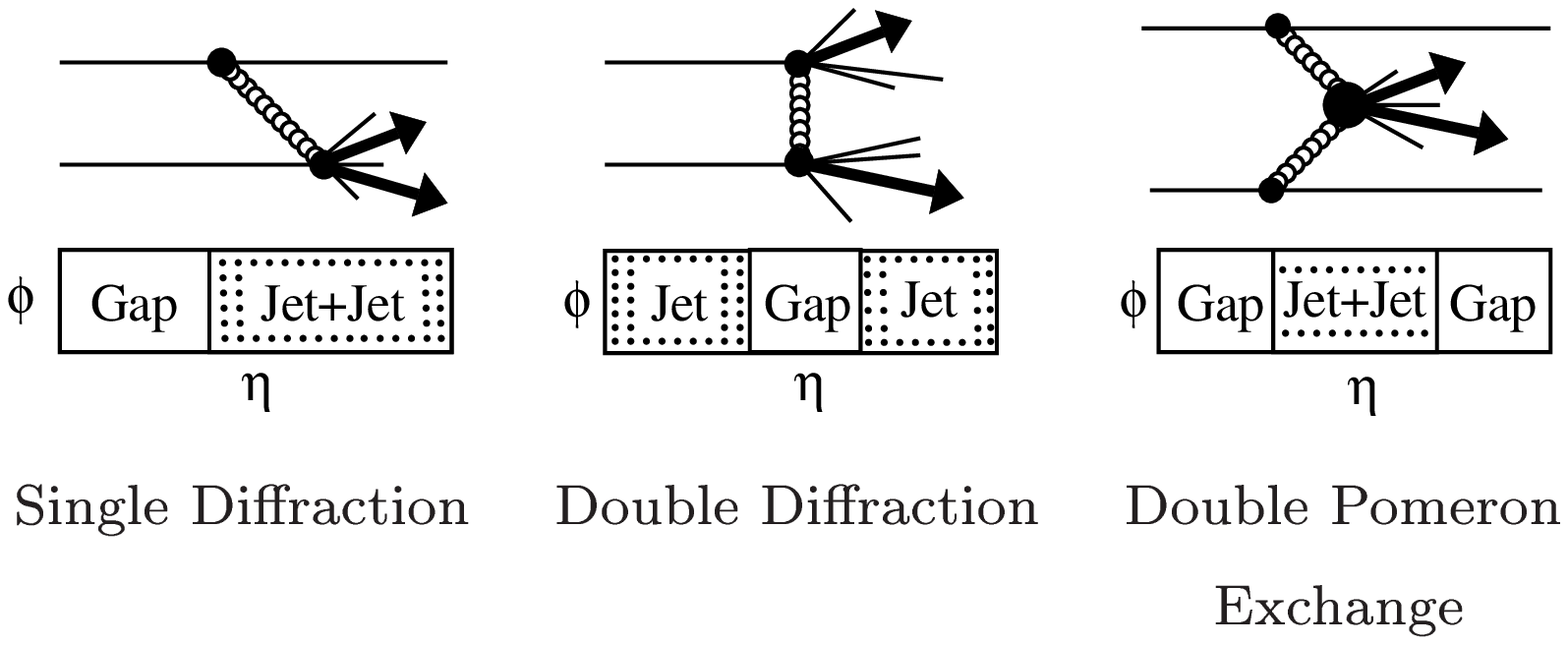,%
        bbllx=58pt,bblly=274pt,bburx=560pt,bbury=510,%
        width=15.0cm,height=7.0cm}}
  \end{picture}
\caption{The three different topologies of hard diffraction at Tevatron.}
\label{fig:tevatron}
\end{center}
\end{figure}

Hard diffractive interactions are studied at the Tevatron by the CDF and
D0 Collaboration, through three different topologies: single diffraction, 
double diffraction and double pomeron exchange, see Fig~\ref{fig:tevatron}.
The single diffraction samples are selected by requiring a 
hard signature ($P_T$ jets, $W$ boson, \jpsi\ meson or a tagged 
$b$ particle), together with the detection of the diffractively
scattered $\bar p$ in the proton spectrometer, or by the presence of a gap 
without activity in the tracker and calorimeter detectors.
The production rate for single diffraction is at the 1 \% level, 
compared to the corresponding non-diffractive process.
Double diffraction process is studied through the production
of two jets separated by a rapidity gap.
The rate for this process was measured at centre of mass energies $\sqrt{s}$ = 630 GeV
and $\sqrt{s}$ = 1800 GeV. The ratio $R_{630/1800}$ is 3.4 $\pm$ 1.2
(2.4 $\pm$ 0.9) for the D0~\cite{d0_ratio} (CDF~\cite{cdf_ratio}) Collaboration.
The decrease of the double diffraction 
process with increasing energy could be explained by the concept of survival 
probability: when energy increases, underlying interactions between the beam particle 
remnants are stronger and can destroy the rapidity gap.
Finally, double pomeron exchange is studied by requiring dijet production
in the central detector and a rapidity gap on both sides of the detector,
or a rapidity gap on one side together with the detection of the diffractively
scattered $\bar p$ in the proton spectrometer. The rate for this process 
is at the level of $10^{-4}$ of the corresponding non-diffractive interactions.

\begin{figure}[bth]
\begin{center}
\setlength{\unitlength}{1cm}
  \begin{picture}(12.0,6.0)
   \put(-.5,0.0){\epsfig{file=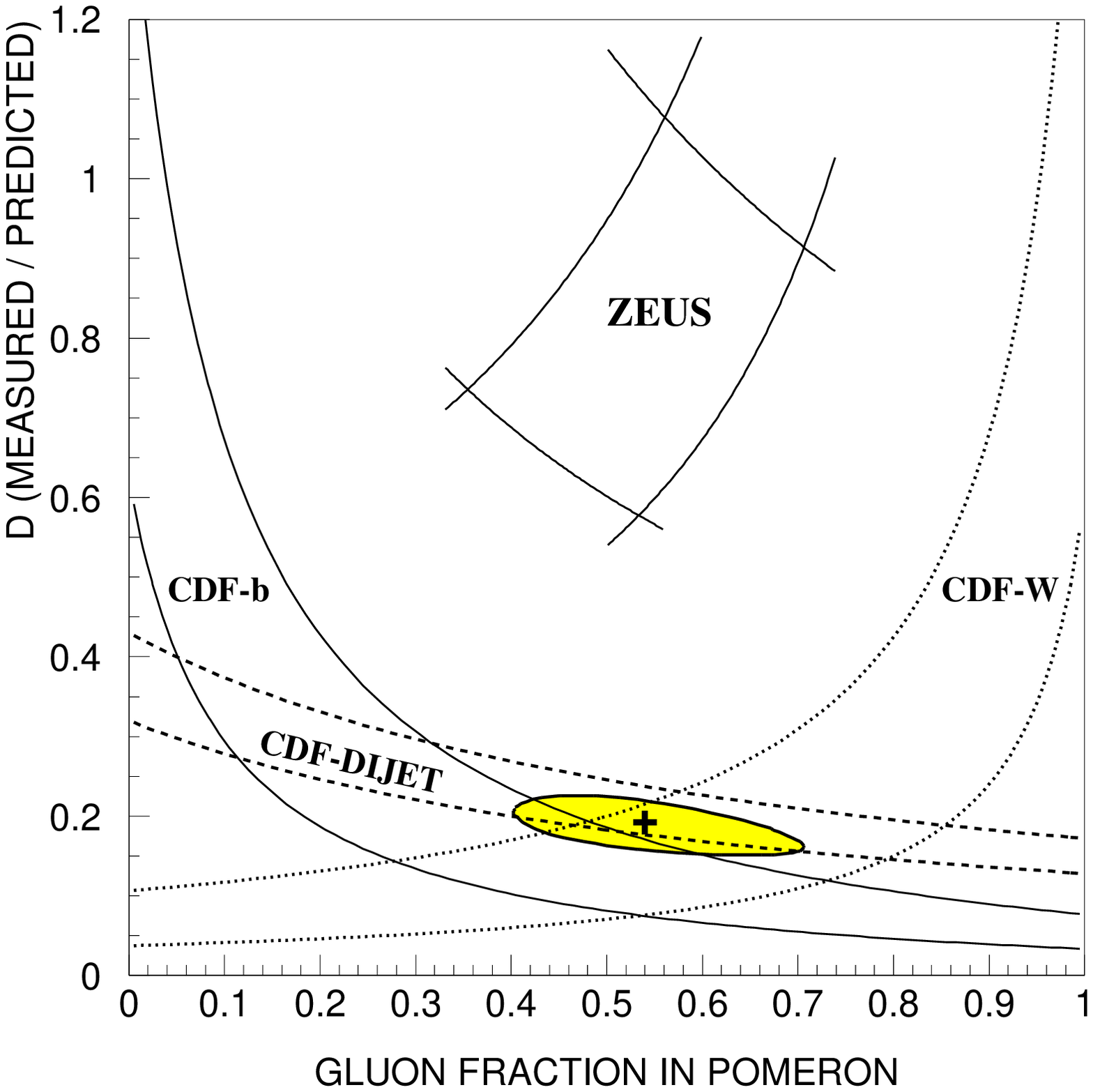,%
        width=6.0cm,height=6.0cm}}
   \put(6.0,0.0){\epsfig{file=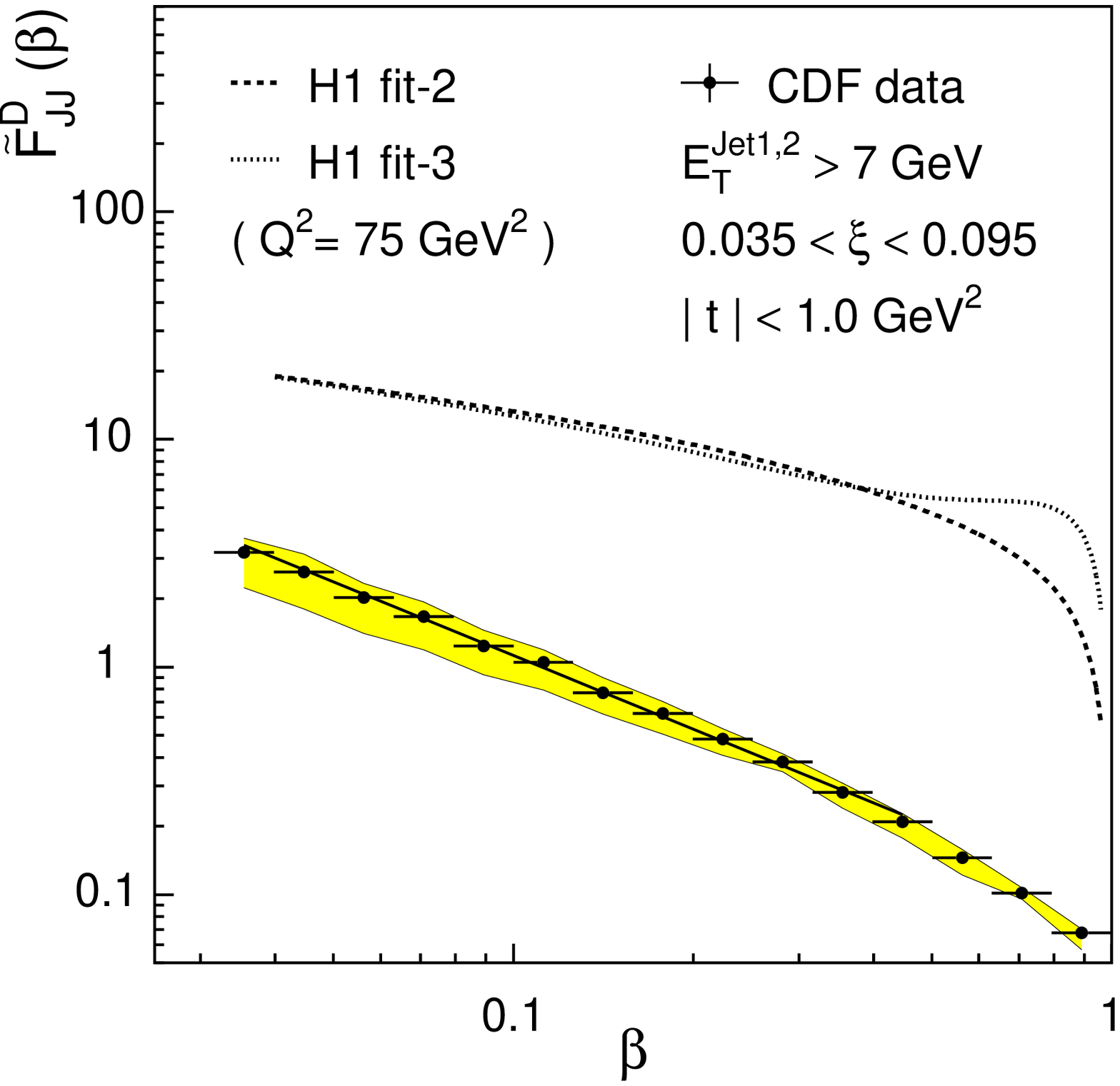,%
        width=6.0cm,height=6.0cm}}
   \put(2.8,-0.6){(a)}
   \put(9.2,-0.6){(b)}
  \end{picture}
\end{center}
\caption{
      (a) Ratio of the measured to predicted diffractive rates as a function of the gluon content
      of the pomeron, for the CDF dijet, $W$ and $b$ production, and for a ZEUS measurement of
      diffractive DIS and diffractive jet
      photoproduction~\protect\cite{cdf_beauty}. The shaded area corresponds to 
      a 1 $\sigma$ contour of a fit to the three CDF results.
      (b) CDF measurement of the diffractive structure function as a function of $\beta$,
      compared with expectations from the pomeron parton densities
      extracted from diffractive
      DIS by the H1 Collaboration~\protect\cite{cdf_dijet}. 
      The systematic uncertainty in the normalisation of the data is $\pm$ 25 \%.}
\label{fig:cdf}
\end{figure}

The CDF Collaboration has determined
the partonic content of the pomeron, taking advantage of the different sensibilities 
of the various processes (dijet, $W$ and $b$ production) to the quark and gluon 
densities~\cite{cdf_beauty}.
The production rates are compared to predictions where a hard partonic content of the 
pomeron was assumed (see Fig.~\ref{fig:cdf}a). 
The gluon density is measured to be 0.55 $\pm$ 0.15, in
agreement with the ZEUS result, but the measured rate at Tevatron is significantly
lower than expected. The $\beta$ dependence of the dijet production cross section, 
measured by the CDF 
Collaboration~\cite{cdf_dijet}, is presented in Fig.~\ref{fig:cdf}b, compared 
to the expected production rate
when using the H1 parton densities for the pomeron. 
The disagreement between the rate of the Tevatron measurement and the 
expectation from HERA indicates a breakdown of factorisation,
which could be understood in terms of survival probability.
%
\section{Exclusive vector particle production at HERA}

Another way to study diffractive interaction is to analyse the
exclusive production of vector meson: $e+p \rightarrow e+p+V$.
For these clean reactions, 
quantitative predictions in perturbative QCD are indeed possible when
a hard scale is present. This scale
can be given by \qsq, $|t|$ or $m_q$, the quark mass.
Most models rely on the fact that, at high energy,
in the proton rest frame, the photon fluctuates into a
$q\bar{q}$ pair a long time before the interaction, and recombines
into a vector meson (VM) a long time after the interaction.
The amplitude ${\cal M}$ then
factories into three terms:
${\cal M} \propto \psi_{\lambda_{V}}^{V \ *}
 \  T_{\lambda_{V} \lambda_{\gamma}} \  \psi^\gamma_{\lambda_{\gamma}}$
where  $T_{\lambda_{V} \lambda_{\gamma}}$ represents the interaction helicity amplitudes
($\lambda_\gamma$ and $\lambda_V$ being the helicities of the
photon and the VM respectively) and $\psi$ represents the wave functions.
In most models, the photon and vector meson $q\bar{q}-p$ interaction is 
described by 2 gluon
exchange. The cross section is then proportional to the square of the
gluon density in the proton:
$ \sigma_{\gamma p} \sim  \alpha_s^2(Q^2) / Q^6 \cdot \left
| xg(x, Q^2) \right| ^2  \label{eq:gluon} $.
The main uncertainties of the models come from the
choice of scale, of the gluon distribution
parameterisation and of the VM wave function (Fermi motion), and from the neglect of
non-diagonal gluon distributions and of higher order corrections.

\begin{figure}[p]
\begin{center}
\setlength{\unitlength}{1cm}
  \begin{picture}(13.0,6.0)
   \put(0.0,0.0){\epsfig{file=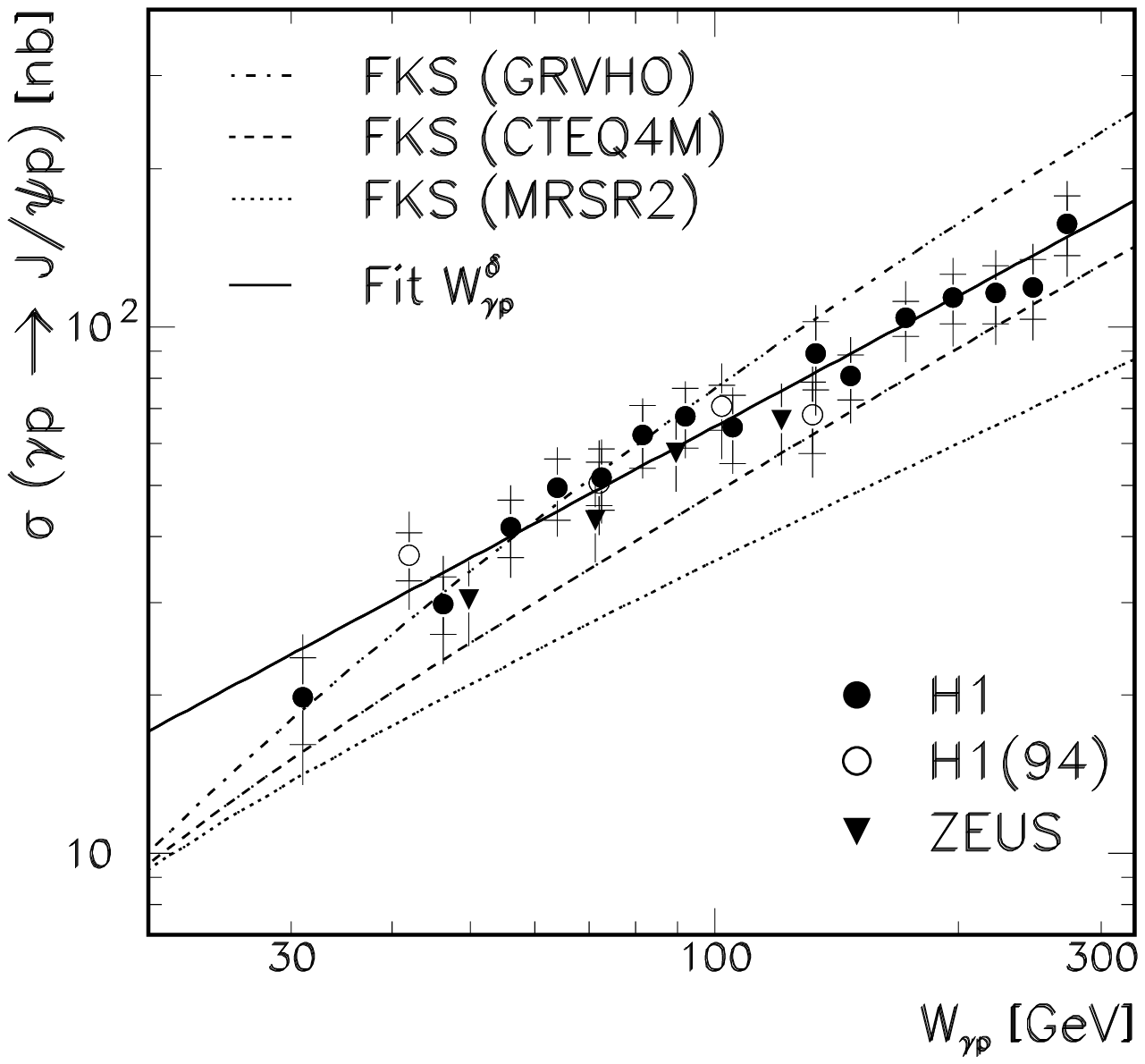,%
           width=6.0cm,height=6.0cm}}
   \put(6.7,-0.1){\epsfig{file=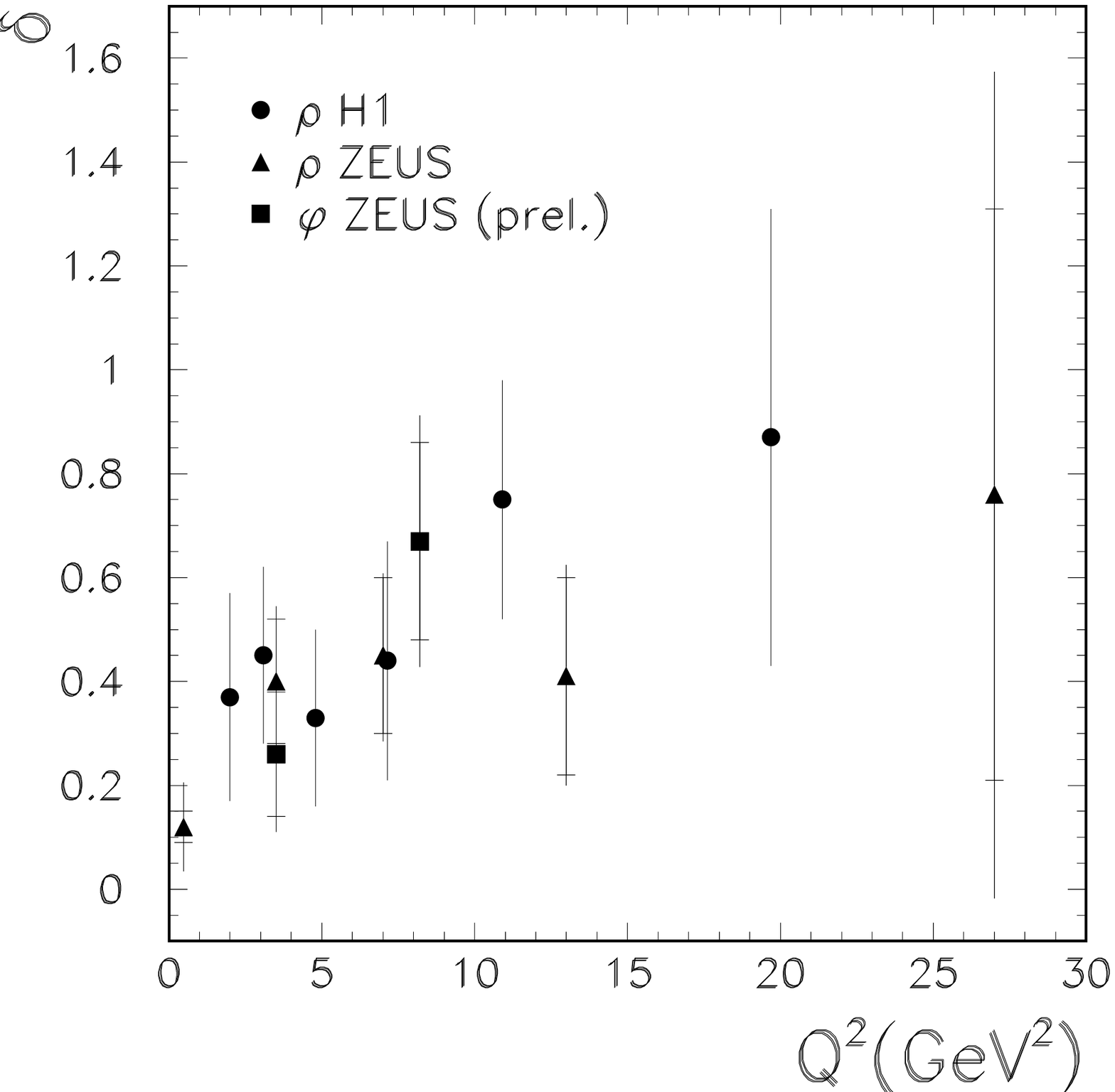,%
           width=6.8cm,height=6.8cm}}
   \put(3.0,-0.2){(a)}
   \put(9.7,-0.2){(b)}
  \end{picture}
\caption{
  a) Cross section $\sigma (\gamma p \rightarrow \jpsi p)$ for \jpsi\ meson
     photoproduction
     as a function of \W. The dotted lines correspond to predictions of 
     a perturbative QCD model~\protect\cite{fks} using different
     gluon density parameterisations and the solid line represents a fit to the
     data using the parameterisation $\sigma \propto W^\delta$, with $\delta$ = 0.83 $\pm$ 0.07;
  b) \qsq\ dependence of the $\delta$ parameter for \rh\ and \ph\ electroproduction.}
  \label{fig:cross}
\end{center}
\end{figure}

Vector meson production has been intensively studied at HERA,
both in photo- (\qsq\ $\simeq$ 0) and electroproduction, for
$\rh, \om, \ph, \jpsi, \psi' {\rm and}~\Upsilon$ mesons.
At high energy, the \rh, \om\ and \ph\ photoproduction cross sections measured by fixed
target experiment and at HERA present a soft energy dependence,
parameterised as $\sigma \propto W^{\delta}$, with $\delta = 0.22$.
In contrast, the \jpsi\ photoproduction cross section, where the mass of the
c quark provides a hard scale in the interaction, presents a much
stronger energy dependence (``hard" behaviour)~\cite{h1_jpsi_ups_gp},
in agreement with the rise of the  gluon density at low $x$ ($x$ $\simeq$ \qsq / $W^2$).
Figure~\ref{fig:cross}a presents the HERA measurement
together with predictions of a
perturbative QCD model~\cite{fks} using three
parameterisations for the gluon density: GRVHO, MRSR2 and CTEQ4M.
The full line corresponds to a fit to the data using the parameterisation
$\sigma \propto W^{\delta}$ with $\delta = 0.83 \pm 0.07$, which is 
in contrast with the value $\delta$ = 0.22 for 
light vector meson photoproduction. 

Another way to look at the hard behaviour is to study
light vector meson production at high \qsq, \qsq\ giving here the scale.
Measurements of the cross section
$\sigma ( \gamma^* p \rightarrow \rho p )$
show an indication for an increasingly stronger energy dependence when
\qsq\ increases~\cite{h1_rho_hq,ze_rho_jpsi_hq}. With the parameterisation
$\sigma  \propto W^{\delta}$, the value of $\delta$ for \rh\ meson production appears 
to reach at high \qsq\ that of \jpsi\ photoproduction (see Fig.~\ref{fig:cross}b).
\begin{figure}[p]
\setlength{\unitlength}{1.0cm}
\begin{center}
 \begin{picture}(14.0,7.0)
    \put(0.0,0.0){\epsfig{file=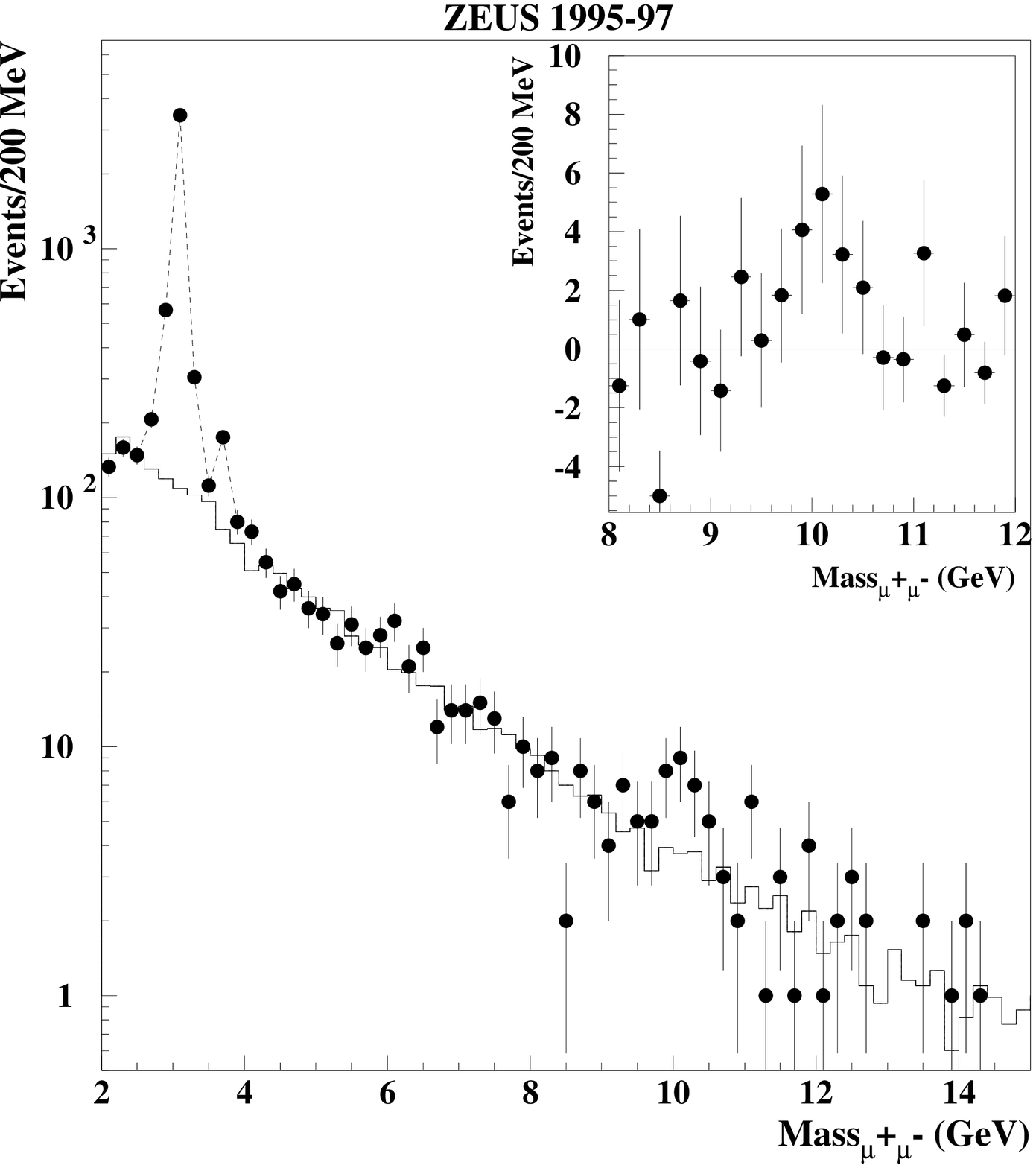,width=7.0cm,height=7.0cm}}
    \put(7.0,-0.3){\epsfig{file=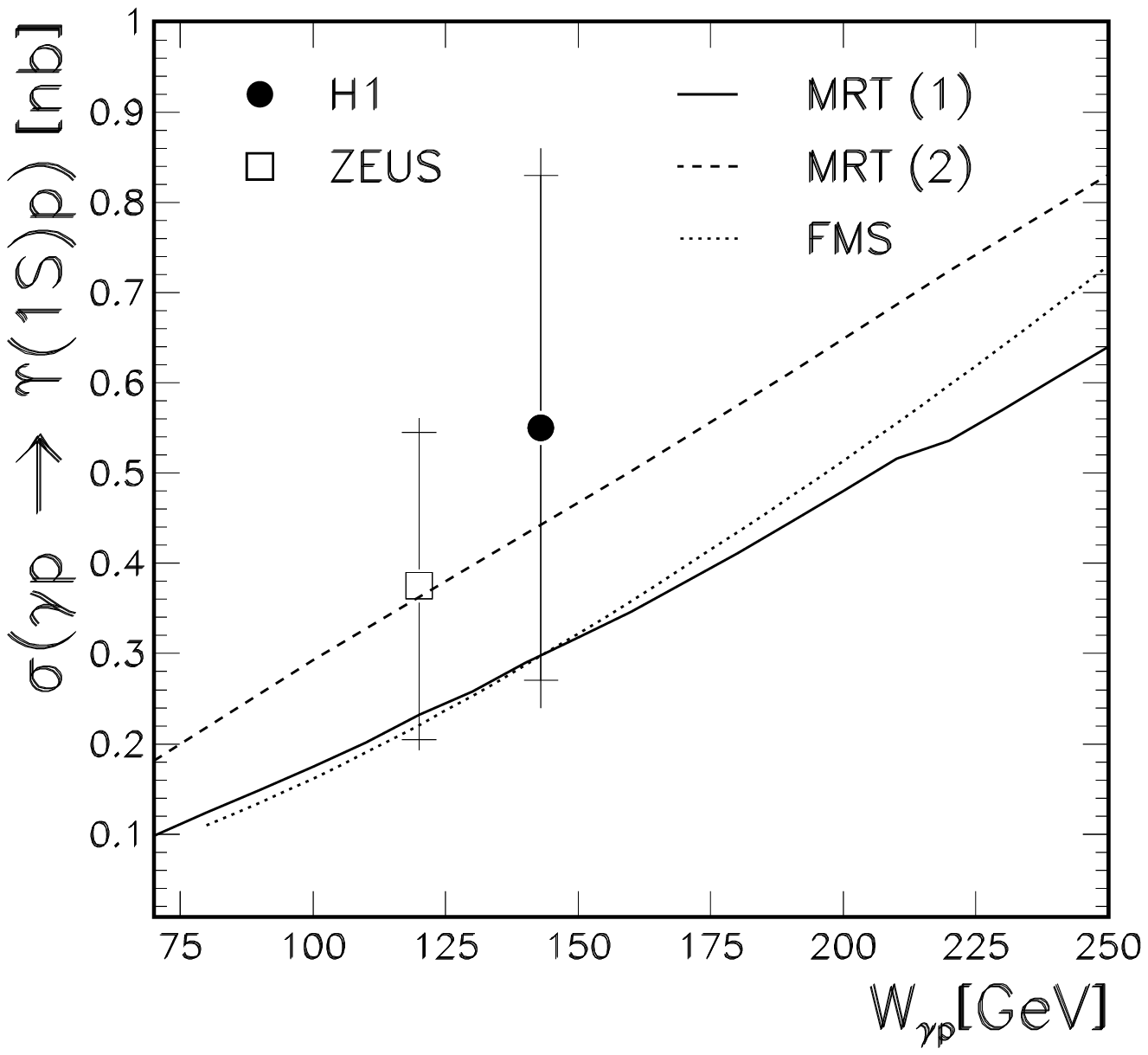,width=6.9cm,height=6.9cm}}
    \put(3.3,-0.3){(a)}
    \put(10.5,-0.3){(b)}
 \end{picture}
\vspace*{0.3cm}
\caption{(a) ZEUS signal for $\Upsilon$ photoproduction~\protect\cite{ze_ups_gp};
    (b) H1~\protect\cite{h1_jpsi_ups_gp} and ZEUS~\protect\cite{ze_ups_gp}
    measurements of the cross section for elastic photoproduction of
    $\Upsilon$ (1S). The curves correspond to predictions of models 
    based on pQCD calculations~\protect\cite{upsilon_calc1,upsilon_calc2}.}
\label{fig:upsilon}
\end{center}
\end{figure}

Signals for $\Upsilon$ production have been
observed recently at HERA~\cite{h1_jpsi_ups_gp,ze_ups_gp}
(see Fig.~\ref{fig:upsilon}a).
The cross section 
$\sigma (\gamma p \rightarrow \Upsilon p) * {\rm BR} (\Upsilon
\rightarrow \mu^+ \mu^-$) for \qsq\ $\simeq$ 0 is measured
to be, respectively, $19.2 \pm 11.0$ and $13.0 \pm 6.6$ pb by the H1 and ZEUS Collaborations;
due to limited statistics the data cannot distinguish between 1S, 2S
and 3S states of the $\Upsilon$ meson. In order to extract  the cross section 
for the production of $\Upsilon$(1S), a production ratio of 70 \% 
is used. The result is shown in Fig.~\ref{fig:upsilon}b, together with recent
pQCD calculations~\cite{upsilon_calc1,upsilon_calc2}. These 
calculations describe well the data after consideration of two effects:
the non vanishing of the real part of the scattering amplitude and the effect of the
non-diagonal parton distributions in the proton, 
which leads to an enhancement of a factor $\simeq$ 5 of the
cross section normalisation. Such effects are found to be more 
important for the production of the
$\Upsilon$ than for the \jpsi\ meson, due to the larger $b$ quark mass.

\begin{figure}[htbp]
\begin{center}
\setlength{\unitlength}{1.0cm}
\begin{picture}(12.0,5.3)
   \put(0.0,0.7){\epsfig{file=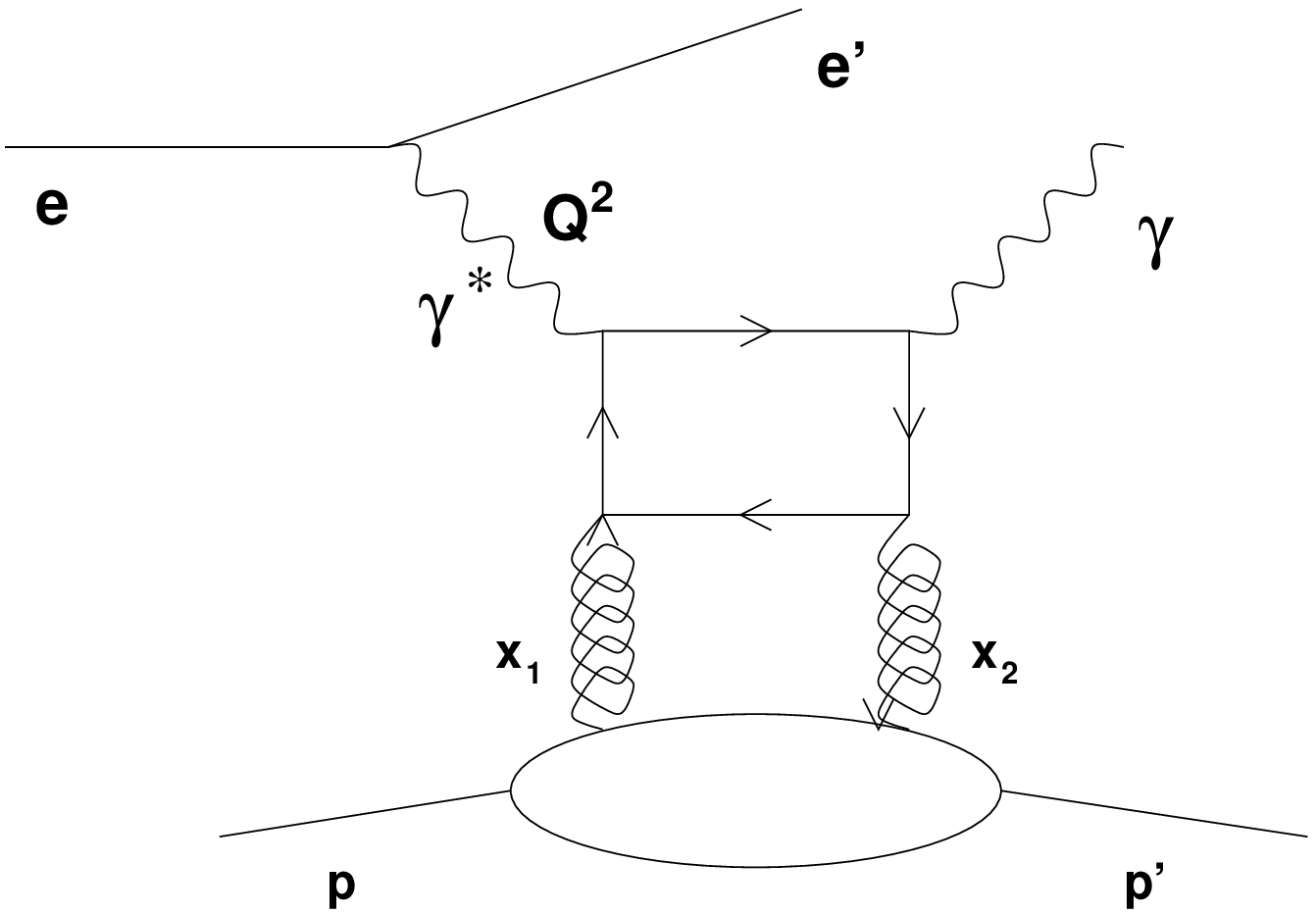,width=5.0cm,height=4.5cm}}
   \put(6.2,0.0){\epsfig{file=dvcsexp_new.ps,%
       bbllx=281.pt,bblly=409.pt,bburx=526.pt,bbury=650.pt,%
       height=5.5cm,width=5.5cm,clip=}}
   \put(2.7,-0.3){(a)}
   \put(9.0,-0.3){(b)}
\end{picture}
\vspace*{0.3cm}
\caption{a) the DVCS process; b) ZEUS measurement of the polar angle distribution of the photon
 candidate, with energy larger than 2 GeV, for $e p \gamma$
 events~\protect\cite{ze_dvcs}.
 The data are the full dots, the predictions for the Bethe-Heitler process
 are the open triangles, and the predictions of a DVCS + Bethe-Heitler
 simulation are the open circles.}
 \label{fig:dvcs}
\end{center}
\end{figure}

The deep virtual compton scattering (DVCS): 
$ e+p \rightarrow e+p+\gamma $ (see Fig.~\ref{fig:dvcs}a)
is a gold-plated process to study pQCD in diffraction.
The reaction is perturbatively calculable 
at high \qsq, as the incoming and the outcoming photon
wave functions and the couplings are known, and no complication
from strong interactions between particle in the final state
appears. It is then an ideal place to study 
non-diagonal parton distributions, thus correlations between 
gluons in the proton.
The main background to the DVCS process is the Bethe-Heitler (QED Compton)
process, which has the same final state but different
phase space.
To extract the DVCS cross section, 
the interference between the two processes has to be taken 
into account.
Events have been selected by the ZEUS Collaboration~\cite{ze_dvcs} 
for \qsq\ $>$ 6 \gevsq, by requesting the presence
of two electromagnetic clusters, corresponding
respectively two the scattered electron and the photon. 
Fig.~\ref{fig:dvcs}b presents the distribution of the polar angle of the photon
cluster.
A clear excess in the data (full points) is observed above the 
Bethe-Heitler background (open triangles). 
The signal is consistent, in shape and in normalisation,
with the prediction of a perturbative QCD calculation computing the
DVCS and the Bethe-Heitler processes, taken into account the interference term (open circles).
%

%

\begin{thebibliography}{99}

\bibitem{f2d3ZEUSlowq2}
 ZEUS Coll., {\em "Measurement of the Diffractive Cross Section in $ep$ Interactions 
  at Low \qsq "}, contributions to the International Europhysics Conference on High Energy Physics 99,
  Tampere, Finland, July 1999.

\bibitem{h1_f2d3}
 H1 Coll., C. Adloff et al., \Journal{\ZPC} {76} {1997} {613}.

\bibitem{h1_dijet}
  H1 Coll., {\em "Diffractive Dijet Electroproduction at HERA"},
  contributions to the International Europhysics Conference on High Energy Physics 99,
  Tampere, Finland, July 1999. 
%
\bibitem{d0_ratio}
  D0 Coll., B. Abbott et al., \Journal{\PLB} {440} {1998} {189}.
%
\bibitem{cdf_ratio}
  CDF Coll., F. Abe et al., \Journal{\PRL} {81} {1998} {5278}.
%
\bibitem{cdf_beauty}
  CDF Coll., T. Affolder et al., \Journal{\PRL} {84} {2000} {232}.
%
\bibitem{cdf_dijet}
  CDF Coll., T. Affolder et al., \Journal{\PRL} {84} {2000} {5043}.
%

%
%
%
%
%
\bibitem{h1_jpsi_ups_gp}
  H1 Coll., C. Adloff et al., {\em "Elastic Photoproduction of \jpsi\ and $\Upsilon$ Mesons 
  at HERA"}, to be publ. in Phys. Lett. B, hep-ex/0003020.
%
\bibitem{fks}
 L. Frankfurt, W. Koepf and M. Strikman, \Journal{\PRD} {57} {1998} {512}.
%
\bibitem {h1_rho_hq}
   H1 Coll., C. Adloff et al., \Journal{\EJC} {13} {2000} {371}.
\bibitem{ze_rho_jpsi_hq}
  ZEUS Coll., J. Breitweg et al., \Journal{\EJC} {6} {1999} {603}.
%
\bibitem{ze_ups_gp}
  ZEUS Coll., J. Breitweg et al., \Journal{\PLB} {437} {1998} {432}.
%
\bibitem {upsilon_calc1}
 L. Frankfurt, M. McDermott and M. Strikman, JHEP {\bf 02} (1999) 002.
%
\bibitem {upsilon_calc2}
 A.D. Martin, M.G. Ryskin and T. Teubner, \Journal{\PLB} {454} {1999} {339}.
%
\bibitem{ze_dvcs}
  ZEUS Coll., {\em "Observation of Deeply Virtual Compton Scattering in $e^+p$ Interactions 
  at HERA"},
  contributions to the International Europhysics Conference on High Energy Physics 99,
  Tampere, Finland, July 1999.
%
%
%
\end{thebibliography}
\end{document}